\documentstyle[prl,aps,preprint,epsf]{revtex}
\begin{document}
\draft
\bibliographystyle{prsty} 
\title{Deformed Fermi Surface Theory of Magneto--Acoustic Anomaly 
in Modulated Quantum Hall Systems Near $\nu=1/2$} 
\author{Nataliya A. Zimbovskaya$^{*}$ and Joseph L. Birman}
\address{Department of Physics, The City College of CUNY, New York,
NY, 10031, USA} 
 
\date{\today} 
\maketitle

\begin{abstract}
 We introduce a new generic model of a deformed Composite Fermion--Fermi 
Surface (CF--FS) for the Fractional Quantum Hall Effect near
$\nu=1/2$ in 
the presence of a periodic density modulation. Our model permits us to 
explain recent surface acoustic wave  observations 
of anisotropic anomalies [1,2] in 
sound velocity and attenuation -- appearance of peaks and anisotropy
-- which 
originate from contributions to the conductivity tensor due to regions 
of the CF--FS which are flattened by  the applied modulation.
The calculated magnetic field and wave 
vector dependence of the CF conductivity,  velocity shift and attenuation
agree with experiments.
              \end{abstract}

PACS numbers 71.10 Pm, 73.40 Hm, 73.20 Dx {}

The integer and fractional quantum Hall effects (IQHE and FQHE)
continue to reveal new and unexpected physics in strongly correlated
2--dimensional electron systems [3]. Recently, particular attention
has been given to FQHE systems at and near half filling of the
lowest Landau level (LLL). According to the present theory at
$\nu = 1/2$ each electron is decorated  by two quantum flux 
tubes, producing  a new fermionic quasiparticle, the composite fermion
(CF) [4]. At $ T=0, $ CFs are distributed inside the Composite 
Fermion--Fermi Surface (CF--FS), which is  assumed to be a circle. When
the filling factor of the LLL is changed from $\nu =1/2$ to $\nu \pm \Delta \nu $, 
the Chern--Simons based theory makes an important prediction. At 
$\Delta\nu = 0$, the two flux
tubes attached to each electron give rise via the ``Chern Simons''
mechanism [3] to an extra (``fictitious'') magnetic field opposite
to and exactly canceling, the applied ${\bf B}$ field. When
$\nu = 1/2 \pm \Delta \nu$ $(\Delta \nu \neq 0)$ the Chern--Simons field
does not cancel the applied field and the CF's move in a non--zero
magnetic field ${\bf B}_{eff}$  which is proportional to $\Delta \nu $. In
order to test the predictions of this theory it is necessary to measure this 
field, and the motion of the carriers in it. A sensitive tool for this 
purpose is surface acoustic wave (SAW) propagation which probes the 
dynamical response of the quantum Hall system to this ${\bf B}_{eff}$ and
gives quantitative information about the carriers [5].

Recently [1,2] anomalous behavior was observed for the SAW velocity
and attenuation near filling factor $\nu = 1/2$ when a periodic
density modulation was applied. Measurements of the velocity shift
$\Delta s/s$ and the attenuation $\Gamma$ in the SAW response
orthogonal to the modulation direction showed an unexpected
effect. The minimum in $\Delta s/s$ at $\nu = 1/2$ which was observed
repeatedly in non--modulated systems [5], was converted to a large
maximum, when the modulation wave vector, and the magnitude of the
external field which produces the modulation, were above some critical
values. On further increase of the magnitude of the density modulation,
the peak in the velocity shift disappeared and was again replaced by a
minimum. For SAW propagation parallel to the direction of density
modulation, no such anomaly was found for the response of the electron
system.

In this paper we will show that a modulation-induced deformation of
the originally circular CF--FS can be at the origin of the observed	
transport anomalies. We assume that exactly at $\nu = 1/2$ the CF-FS
is a circle, with radius $p_{F} = (4 \pi n \hbar^{2})^{1/2} $, where
$n$ is the electron density. In the presence of the grating modulation
the CF--FS circle is ``flattened'' in the neighborhood of two special
points where the curvature  vanishes. Such small, locally ``flat'' 
regions can under certain conditions, play a disproportionately important 
role in determining the magneto--conductivity response due to the 
unusually large density of quasiparticle states there. The response is 
very sensitive to local changes of FS topology 
as we show below by introducing a concrete model which  permits us to obtain
analytical expressions for $\Delta s/s$ and $\Gamma$. Using appropriate
parameters we obtain semiquantative
agreement with experiment. The model also explains the orthogonality of
response and predicts its wave-vector dependence. We conjecture  below 
that the reason for the reported  disappearance of peaks at the highest 
modulation  is related to additional topological change in the CF--FS.

As a first step, assume the periodic modulation in the 
y-direction introduces a single Fourier  component of potential $V_{g}$ 
into a  ``nearly--free'' particle CF model. The resulting dispersion 
relation is: 
   \begin{equation}
E({\bf p})= \frac{p_x^2}{2m^*} + 
\frac{p_y^{*2}}{2m^*} + 
\frac{(\hbar g)^2}{8m^*} - 
\sqrt{\left( \frac{\hbar g p_y^*}{2m^*} \right)^2 + V_{g}^{2}},
                         \end{equation}
with $p_{y}^{*}=p_{y}-\hbar g/2$, $m^{*}$ is the CF effective
mass. The curvature of the 2--D CF--FS can also be directly calculated as:
                 \begin{equation}
\kappa = \left [2v_{x} v_{y} \frac {\partial v_x}{\partial p_y}
- v_x^2 \frac {\partial v_y}{\partial p_y} - v_{y}^{2}
 \frac {\partial v_x}{\partial p_x} \right]  \left / v^{3}, \right.
                \end{equation}
with $v=\sqrt{v_{x}^{2}+v_{y}^{2}}$. The curvature $\kappa$ tends to
zero when $ p_x \rightarrow \pm p_F \sqrt {V_g/ E_F} $. The
importance of this is that near to these points on the CF--FS the CF
velocities are nearly parallel to the $y$ direction. When
$ ql\gg 1$ ($q$ is the SAW wave vector; $l$ is the CF mean
free path) these parts of the CF--FS make the major contribution to the
velocity shift $ \Delta s/s $ and attenuation $ \Gamma $ of the SAW
propagating in the $x$ 
direction. Near these zero curvature points we will use asymptotic
expressions for Eq.(1). Determining $(p_{x}^{*},p_{y}^{*})$ by
$p_{x}^{*}=\eta p_{F} $, 
$ \displaystyle {p_{y}^{*}=p_{F} \left (1-\frac{1}{\sqrt{2}}\eta^{2}
\right )} $, where
$\eta=\sqrt{V_{g}/E_{F}}$, $E_{F}=p_{F}^{2}/2m^{*}$, we can
expand the variable $p_{y}$ in powers of $(p_{x}-p_{x}^{*})$, and
keep the lowest order terms  in the expansion. We obtain:
               \begin{equation}
p_{y}-p_{y}^{*}=-\eta(p_{x}-p_{x}^{*})-
\frac{2}{\eta^{4}}\frac{(p_{x}-p_{x}^{3})^{3}}{p_{F}^{2}}.
                    \end{equation}
Near $p_{x}^{*}$, where $(|p_{x}-p_{x}^{*}|<\eta^{2}p_{F})$ the first term on
the right side of Eq.(3) is small compared to the second one and can
be omitted.

Hence near $p_{x}^{*}$ we have:
                       \begin{equation}
E({\bf p})=\frac{4}{\eta^{4}}\frac{p_{F}^{2}}{2m^{*}}
\left(\frac{p_{x}-p_{x}^{*}}{p_{F}}\right)^{3}+\frac{p_{y}^{2}}{2m^{*}}.
                                   \end{equation}
Using experimental values from [1] we find $V_{g}\sim
10^{-2}ev$, $n\sim 10^{12}cm^{-2}$. Hence $V_{g}$ is not small compared
to the Fermi energy and the local flattening of the CF--FS can be
quite significant. 

To analyze the contribution to the conductivity from these flattened
parts we  generalize the expression for $E({\bf p})$ and define our model as:
          \begin{equation}
E({\bf p}) = \frac{p_{0}^{2}}{2m_{1}} \left |
\frac{p_{x}}{p_{0}} \right |^{\gamma} + \frac{p_{y}^{2}}{2m_{2}}, 
                                \end{equation}
where $p_{0}$ is a constant with the dimension of momentum, the
$m_{i}$ are effective masses, and $\gamma$ is a dimensionless
parameter which will determine the shape of the CF--FS . We shall 
take $\gamma > 1 $ to avoid singularities in the CF velocity.
When $\gamma>2$ the 2--D CF--FS looks like an ellipse 
flattened near the vertices $(0, \pm p_{0})$. Near these points the 
curvature is:
                    \begin{equation}	
\kappa = -
\frac{\gamma(\gamma-1)}{2p_{0}\sqrt{m_{2}/m_{1}}}
\left |\frac{p_x}{p_0} \right |^{\gamma-2}
          \end{equation} 
and, $ \kappa \rightarrow  0$ at $p_{x} \rightarrow 0$. The CF--FS will be
the flatter
at $(0, \pm p_{0})$, the larger is the parameter $\gamma$. We can now 
calculate the desired responses to the SAW, using Eq.(5).

In a GaAs heterostructure with a 2-D electron gas subject to a 
travelling SAW, piezoelectric coupling 
produces a longitudinal electric field which interacts with the
electron gas. Taking the SAW wave vector as $(q,0,0)$ we obtain that
the resulting velocity shift $\Delta s/s$ and SAW attenuation rate
$\Gamma$ are given by the following expressions [6]
                        \begin{eqnarray}
\Delta s/s = [\alpha^2/2]
\Re (1 + i\sigma_{xx}/\sigma_{m} )^{-1}, \\   
\Gamma=-q(\alpha^{2})/2\Im(1+i\sigma_{xx}/\sigma_{m})^{-1}.
\end{eqnarray}
In these equations, $\omega=sq$ is the SAW frequency, $\alpha$ is the 
piezoelectric coupling constant, $\sigma_{m}=\epsilon s/(2\pi)$ with 
$\epsilon$ an effective dielectric constant of the medium,
$\sigma_{xx}$ is the  $xx$ component of the electronic conductivity
tensor; real and imaginary parts are indicated. In order to proceed we now
need to establish some
preliminary results.
We use the semi--classical CF theory [4] in which the CF quasiparticles 
have charge $e$, and finite mass $ m^*. $  However, as described below, a 
particular variant of the solution of the Boltzmann equation was needed 
for the present work. In semiclassical CF theory   the electron 
resistivity tensor $\rho$ at finite $({\bf q},\omega)$ is the sum of a CF 
term and a term originating in the magnetic field of  the
Chern-Simons (CS) vector potential. The CS part has only 
off--diagonal elements,
                   \begin{equation}
(\rho^{CS})_{xy}=-(\rho^{CS})_{yx} = 4 \pi \hbar/e^2.
                                               \end{equation}
In a strong magnetic field we have $\rho_{xy}\gg\rho_{xx}$, $\rho_{yy}$,
and hence we can use the approximation:
                         \begin{equation}
\sigma_{xx}({\bf q})=e^{4}/[(4\pi\hbar)^2\tilde{\sigma}_{yy}(\bf{q})],
                       \end{equation}
where $\tilde{\sigma}=(\rho^{CF})^{-1}$ is the CF 
conductivity. 

We shall calculate the CF conductivity tensor at $\nu$ close to $1/2$,
using the resulting non-zero effective magnetic field which contains a
spatially nonuniform contribution $\Delta {\bf B} \exp(igy)$ associated
with
the electron density modulation $\Delta n(y)$ as: $ \Delta
B(y)= - (4\pi\hbar c/e)\Delta n(y)$. Consequently, we  assume we can
replace
the initial system of CFs with a system of quasiparticles containing 
$n+<\Delta n>$ negatively 
charged quasielectrons (fermions) and $<\Delta n>$ positively charged
quasiholes (fermions) per unit area. Here $<\Delta n>$ is equal to
the root mean-square value of $<\Delta n(r)>$, and corresponds to
the additional average fictitious magnetic field $<\Delta {\bf B}>$. We
assume
that we can consider the response of this two component system in the
uniform effective magnetic field
 $ {\bf B}_{eff}^{*} = {\bf B}_{eff}+<\Delta {\bf B}>$,
instead of the response of the initial one-component CF system in the
nonuniform effective magnetic field. 

To evaluate the quasielectron contribution to the CF conductivity
$\tilde{\sigma}^{e}_{\alpha\beta}({\bf q})$ so that we can pass
smoothly to the $B_{eff}\rightarrow 0$ limit for a flattened CF--FS, we
begin with the expression obtained from solution of the linearized
Boltzmann equation in the presence of the magnetic field, assuming a
relaxation time $\tau. $ This is:
                      \begin{eqnarray}
\tilde{\sigma}_{\alpha\beta}^{e}(\nu) =
 \frac{e^{2}m_{c}}{(2\pi\hbar)^{2}}\frac{1}{\Omega}
\int \limits_{0}^{2 \pi}d\psi \left \{ \exp \left
[-\frac{iq}{\Omega} \int \limits_{0}^{\psi}V_{x}
(\psi^{\prime\prime})d\psi^{\prime\prime}\right]V_{\alpha}(\psi) 
       \right.      \times \nonumber \\ 
\times     \left.
\int \limits_{-\infty}^{\psi} \exp
\left [\frac{iq}{\Omega} \int
\limits_{0}^{\psi'} V_{x}(\psi^{\prime})d\psi^{\prime}
+ \frac{1}{\Omega\tau}(\psi^{\prime} -\psi)\right ]
V_{\beta}(\psi^{\prime})d\psi^{\prime} \right \}.
                  \end{eqnarray}
Here $V_{\alpha,\beta} $ are the quasielectron velocity components
$(\alpha,\beta=x,y)$; $\Omega=|e|B_{eff}^{*}/m_{c}c$ is their
cyclotron frequency; $\psi$ is the angular coordinate on the
quasielectron cyclotron orbit, ($\psi=\Omega\theta$; $\theta$ is the
time of the quasielectron motion along the cyclotron orbit). We have
taken $\omega\tau \ll 1$. We proceed [7] as follows. Express the
velocity components $V_{\beta}(\psi^{\prime})$ as  Fourier series:
                        \begin{equation}
V_{\beta}(\psi^{\prime})=\sum_{k} V_{k\beta} \exp(ik\psi^{\prime}).
            \end{equation}
Introducing a new variable $\eta$:
                  \begin{equation}
\eta \equiv \left (\frac{1}{\tau} + ik\Omega+iqV_{x}(\psi)\right) 
\tilde{\theta}+iq\int \limits_{0}^{\tilde{\theta}}
[V_{x}(\psi+\Omega\theta^{\prime})-V_{x}(\psi)]d\theta^{\prime}
                                \end{equation}
and substituting (11) and (12) into (10) we obtain:
                  \begin{equation}
\tilde{\sigma}_{\alpha\beta}^{e}(\nu) = 
\frac{e^{2}m_c\tau}{(2\pi\hbar)^{2}}
\sum_{k}V_{k\beta} \int \limits_{-\infty}^{0}e^{\eta} d\eta
\int \limits_{0}^{2\pi}\frac{V_{\alpha}(\psi)
\exp(ik\psi)d\psi}{1+ik\Omega\tau +
iqV_{x}(\psi+\tilde{\theta} (\eta)\Omega)\tau}\, .
                             \end{equation}
To proceed we can transform the integral over $\psi$ in (14) to an
integral over the CF--FS. Reexpressing the element of integration as
$ m_c d\psi = d \lambda /|v| \; \; (d\lambda $ is the element of
length along the Fermi Arc) and $ m_c $ will be replaced by a  suitable
combination of $ m_1, m_2 $ of our model (5); e.g. for an ellipse $ m_c =
\sqrt {m_1 m_2}. $ We can now parameterize the dispersion equation of our
model (5) as follows:
                   \begin{equation}
p_{x}=\pm p_{0}|\cos t|^{2/\gamma}; \qquad \qquad
p_{y}=p_{0}\sqrt{m_{2}/m_{1}} \, \sin t,
           \end{equation}
where $0\leq t\leq2\pi$, and the $+$ and $-$ signs are chosen
corresponding to normal domains of positive and negative values of the 
cosine. Where $ql \gg 1, $ the leading term in the resulting formula
originates from parts of the CF--FS where $v_{x}\approx 0$. Expanding
it in powers of $(ql)^{-1}$ and keeping the main term in the expansion
we obtain:
                           \begin{equation}
\tilde{\sigma}_{yy}^{e}(\nu) = 
\frac{b}{2} \frac{e^2 p_0}{4 \pi\hbar^2} \frac{l}{ql^\mu} 
(S_{+\mu}(\Omega\tau)+S_{-\mu}(\Omega\tau))
                                           \end{equation}
where: $\displaystyle {S_{\pm\mu}(\Omega\tau) = \int\limits_{-\infty}^{0}
e^\eta(1\mp i\Omega\tau(1\pm\eta\delta_{0}))^{\mu-1}d\eta} $ and
$\delta_{0}$ is a small dimensionless constant of the order of
$\omega\tau$.
Here for convenience we introduced $\mu=1/(\gamma-1)$ which is a 
dimensionless parameter $(0\not=\mu\leq 1)$, with $\mu=1$, or $\gamma=2$
corresponding to the case that the CF--FS is an ellipse.
In these variables, the CF mean-free-path $\ell$ is equal to:
                    \begin{equation}
\ell = \frac{\mu + 1}{2 \mu} \frac{p_0 \tau}{m_1}.
                                  \end{equation}
Passing to the limit $B_{eff}=0$ we have:
                                  \begin{equation}
\tilde{\sigma}^e_{yy} \left (\nu = \frac{1}{2} \right) =
\frac{be^2 p_0}{4\pi\hbar^2} \frac{\ell}{(q \ell)^\mu} \, . 
                                    \end{equation}
In this equation $ b =
4\mu^{2}/(\mu+1)\sqrt{m_{1}/m_{2}}[\sin(\pi\mu/2)]^{-1}$.
This expression eqn.(18)  predicts that  measuring the $q$-dependence of
the conductivity 
exactly at $B_{eff}^{*}=0$ ($\nu=1/2$) can give the deformation
parameter $\mu$. When the CF--FS is an undeformed circle
( $m_{1} = m_{2} = m^*) $ then $b=2$ and the result is identical to the 
corresponding result obtained in [4]. It is worth emphasizing that
under 
the condition that the flattening of the CF-FS is strong, with  $\gamma\gg 
1$, the quantity $\mu\approx 0$ and  the CF conductivity will be enhanced 
compared to the circular case, and it will be effectively independent
of q (See eqn.(16)). Independence of $q$ has been found experimentally
[1]. For small $\Omega\tau, \; (\Omega \tau \omega\tau < 1)$ one can
expand the functions $ S_{\pm \mu}(\Omega\tau) \; (\mu \neq 1) $ in
powers
of $\delta_{0}\Omega\tau$:
                         \begin{equation}
S_{\pm \mu}(\Omega\tau) = 
(1\mp\Omega\tau)^{\mu-1}\left[1+\sum_{r=1}^{\infty}
\frac{(1-\mu)(2-\mu)...(r-\mu)}{(1\mp
i\Omega\tau)^{r}}(i\delta_{0}\Omega\tau)^{r}\right].
                                                  \end{equation}
Keeping the terms larger than $(\Omega\tau)^3$ one has:
                         \begin{equation}
\tilde{\sigma}_{yy}^{(e)} = 
\tilde{\sigma}_{yy}^{e} \left (\nu = \frac{1}{2} \right ) 
[1-a^{2}(\Omega\tau)^2 + i \xi \Omega\tau] .
                          \end{equation}
Here $a^{2}=((1-\mu)(2-\mu)/2)(1+2\delta_{0}^{2})$ and
$ \xi =(1-\mu)\delta_{0}$ are positive constants. For sufficiently small
values of the parameter $\mu$ (significant flattening of the effective
parts of the CF--FS) the constant $a^{2}$ is of the order of unity and
the constant $ \xi $ is small compared to unity, because of the small
factor $\delta_{0}$. 

In the experiments [1,2]
the quasihole density $<\Delta n>$ and the corresponding 
Fermi momentum is small compared to these for the quasielectrons. Therefore
the quasihole contribution to the CF conductivity can be neglected.
Substituting the result (16) into (10), we can obtain the expression
for the electron conductivity component $\sigma_{xx}$. The using
(7),(8) we have:
                 \begin{equation}
\frac{\Delta s}{s} = \frac{\alpha^{2}}{2}
\frac{1+ \xi\Omega\tau\bar{\sigma}}{1+\bar{\sigma}^2}  
\left(1-\frac{2\xi\Omega\tau\bar{\sigma}}{1+\bar{\sigma}^{2}} - 
\frac{\bar{\sigma}^{2}}{1+\bar{\sigma}^{2}}
(2a^{2}-\xi^{2})(\Omega\tau)^{2}\right) ;
                                          \end{equation}
                         \begin{equation}
\Gamma = q\frac{\alpha^{2}}{2} 
\frac{\bar{\sigma}^{2}}{1+\bar{\sigma}^{2}}
\left(1-\frac{2 \xi\Omega\tau\bar{\sigma}}{1+\bar{\sigma}^{2}} 
-\frac{a^{2}\bar{\sigma}^{2}}{1+\bar{\sigma}^{2}}(\Omega\tau)^{2}\right) .
                                     \end{equation}
Here $\bar{\sigma}=\sigma_{xx}(\nu=1/2)/\sigma_{m}. $
Expression (21) and (22) are the new results of our theory. They predict
peaks both in the SAW attenuation and velocity shift at $\nu=1/2$. the
peaks arise due to distortion of the CF--FS in the presence of the
density modulation. When the CF--FS flattening is strong ($\mu\ll 1$)
the magnitude of the peak of the velocity shift is practically
independent of the SAW wave vector $q$. Also these anomalies are not
sensitive to any relation between $q$ and the density modulation wave
vector $g$. As was observed repeatedly [1,2] the peaks appear when  the 
magnitude of the modulating potential and its
wave vector are sufficiently large. These quantities $V_{g}$ and $g$
determine the character and amount of distortion of the CF--FS.

Our model allows us to obtain the dependence of the conductivity on
filling factor $\nu$ for the undistorted (circle) CF--FS, when
$\mu=1$. In that case, the main term in the expansion of the CF
conductivity in inverse powers of $ ql $ is independent of the magnetic
field. When we take  into account the next term of the expansion we 
arrive at the following expression after a lengthy, but
straightforward calculation [10]:
                     \begin{equation}
\tilde{\sigma}_{yy}^{e}(\nu) =
\frac{2e^{2}p_{F}^{2}\tau}{(2\pi\hbar)^{2}m^*}
\int \limits_{-\infty}^{0} e^{\eta} 
\left(\frac{\pi}{ql} + 2\Omega\tau
\frac{\ln(ql)}{(ql)^2} + \Omega\tau f(\eta) O(ql)^{-1} \right) d\eta
                                                           \end{equation}
where $ O $ means ``order of ``.

Now, noting that the first two terms are independent of $\eta $, and
retaining only them in doing the integral, we obtain the CF
conductivity tensor component: 
                        \begin{equation}
\tilde{\sigma}_{yy}(\nu) = 
\tilde {\sigma}_{yy} \left (\nu = \frac{1}{2} \right) 
\left [1 + \Omega \tau \frac {2 \ln(ql)}{\pi ql} \right] .
                                                    \end{equation}
Expression (23) describes the CF conductivity increasing as $B_{eff}$
increases. This corresponds to the minimum in the SAW velocity shift
at $\mu=1/2$. This minimum was observed repeatedly in non-modulated FQHE
systems [5].

We now suggest  an explanation for the observed disappearance of the SAW peak
in $\Delta s/s$ when the magnitude of density modulation was at
highest measured values. In metals it is known [8,9] that external
factors, as well as changes in electron density can cause changes in FS
topology. Such changes are sensitively reflected in the response
functions. We suggest this can occur in the CF case. A topological
change of  the CF--FS connectivity can be caused by increased magnitude
of modulating field and correspondingly increased quasiparticle
density $n+\langle {\Delta n}\rangle$. Changing the CF--FS connectivity can 
lead to the disappearance of the flattening of the effective parts of the
CF--FS. In this case the anomalous maximum in the magnetic field
dependence of $\Delta s/s$ will be replaced by minimum. Thus assuming
the relevance of the electron topological transition, we can explain
the disappearance of the peak in the SAW velocity shift under increase
of the modulation strength. Additional experimental consequences of our 
model and more details of the theory will be presented elsewhere [10].

Recently, several other theoretical papers have discussed  [11,12]
the 
experiments on density modulated systems near $\nu = 1/2, $
although 
our explicit deformed CF--FS model is 
new, to our knowledge. A point of contact between our work and that  
of [11] may be their assertion of anisotropic resistivity due to the 
spatially averaged current and electric field in the presence of 
periodically modulated quasiparticle density [see eqn (2) of ref 11].
This assertion seems implicitly to correspond to our deformed CF--FS; the 
two approaches would then be equivalent when $\Delta n \ll n$. 
However, in [11,12] it was assumed 
that the wavelength of the density modulation is small compared to the SAW
wavelength ($g \gg q$), and also that the interaction energy of the  CF
with the  modulating field is small compared to the Fermi energy. According
to our estimate as presented above, these assumptions do not
correspond to the reported experiments.  

We again remark that our work is based on the charged CF picture for FQHE,
for example as derived at $\nu = 1/2 $ in ref. [4] from a Chern--Simons 
approach. An alternate picture for the FQHE also derived  from a 
Chern--Simons approach, gives the quasiparticles at $\nu = 1/2$ as neutral 
dipolar objects, with the Hall current being carried by a set of collective
magneto--plasmon oscillators. To our knowledge, a magneto--transport
theory based 
on this second picture does not exist at present, so we are not able to 
compare our results with any derived from that picture.

We thank Dr.R.L.Willett for discussions, Mr. Meng Lu and Dr.Gregory
Zimbovsky for help with the
manuscript. This work was supported in part by a grant from the 
National Research Council COBASE Program. 

------------------------------------------------------------------

* Permanent address (after October, 1998): Urals State Mining and
Geological Academy, Ekaterinburg, 620144, Russia

1. R.L.Willet,  K.W.West and L.N.Pfeiffer, Phys. Rev. Lett., {\bf 78}, 
4478 (1997).

2. J.H.Smet, K. von Klitzing, D.Weiss and W.Wegscheider, Phys. Rev. Lett.,
{\bf 80}, 4538 (1998).

3. ''The Quantum Hall Effect'' ed R.E.Prange and S.M.Girvin (Springer,
NY 1987); ''Perspectives in Quantum Hall Effect''
ed. S. das Sarma and A.Pinczuk (J.Wiley, 1997).

4. B.I.Halperin, P.A.Lee and N.Read, Phys. Rev. B {\bf 47}, 7312 (1993); 
S.N.Simon and B.I.Halperin, Phys. Rev. B {\bf 48}, 17368 (1993);
A.Stern and B.I.Halperin, Phys. Rev. B {\bf 52}, 5840 (1995).

5. R.L.Willet. Advances of Physics, {\bf 46}, 447, (1997).

6. K.A.Inbergrigsten, J. Appl. Phys. {\bf 40}, 2681 (1969);
P.Bierbaum, Appl. Phys. Lett. {\bf 21}, 595 (1972).

7. N.A.Zimbovskaya, V.I.Okulov, A.Yu.Romanov and V.P.Silin,
Fiz. Met. Metalloved., {\bf 62}, 1095 (1986) (in Russian);
N.A.Zimbovskaya, Fiz. Nizk. Temp. {\bf 20}, 441 (1994); [Sov. Low
Temperature Physics, {\bf 20}, 324 (1994)].
N.A.Zimbovskaya, ''Local Geometry of the Fermi Surface 
and High Frequency Phenomena in Metals.'' ''Nauka'', Ekaterinburg, 1996
(in Russian).

8. I.M.Lifshitz, Zh. Eksp. Teor. Fiz. {\bf 38}, 1569 (1960)
[Sov. Phys JETP {\bf 11}, 1130 (1960)].

9. Ya.M.Blanter, M.I.Kaganov, A.V.Pantsulava and A.A.Varlamov, Physics
Reports, {\bf 245}, 159 (1994).

10. N.A.Zimbovskaya and J.L.Birman (in preparation).

11. Felix v. Oppen, Ady Stern and B.I.Halperin,  Phys. Rev. Lett., 
{\bf 80}, 4494 (1998).

12. A.D.Mirlin, P.Wolfle, Y.Levinson and O.Entin-Wohlman,
cond-mat/9802140.

\end{document}